\documentclass[%
 reprint,
 amsmath,amssymb,
 aps,
]{revtex4-2}

\usepackage{color}
\usepackage{comment}
\usepackage{graphicx}
\usepackage{dcolumn}
\usepackage{bm}
\usepackage{algorithm}
\usepackage{algorithmic}
\usepackage{bbold}
\usepackage{soul}
\usepackage{graphics}

\usepackage{inputenc}
\inputencoding{utf8}
\makeatother

\usepackage{cellspace}
\usepackage{makecell}
\usepackage{hyperref}

\begin{document}

\preprint{APS/123-QED}

\title{Learning deterministic hydrodynamic equations from stochastic active particle dynamics}

\author{Suryanarayana Maddu$^{1,2,3,4}$, Quentin Vagne$^{2,3,6,7}$, Ivo F.~Sbalzarini$^{1,2,3,4,5}$} 
\affiliation{ $^{1}$ Technische Universit\"{a}t Dresden, Faculty of Computer Science, 01069 Dresden, Germany}
\affiliation{$^{2}$ Max Planck Institute of Molecular Cell Biology and Genetics, 01307 Dresden, Germany}
\affiliation{ $^{3}$ Center for Systems Biology Dresden, 01307 Dresden, Germany} 
\affiliation{ $^{4}$ Center for Scalable Data Analytics and Artificial Intelligence ScaDS.AI, Dresden/Leipzig, Germany}
\affiliation{ $^{5}$ Cluster of Excellence Physics of Life, TU Dresden, Germany}
\affiliation{ $^{6}$ Max Planck Institute for the Physics of Complex Systems, 01187 Dresden, Germany}
\affiliation{ $^{7}$ Department of Genetics and Evolution, University of Geneva, Switzerland}
\date{\today}

\begin{abstract}
We present a principled data-driven strategy for learning deterministic hydrodynamic models directly from stochastic non-equilibrium active particle trajectories. 
We apply our method to learning a hydrodynamic model for the propagating density lanes observed in self-propelled particle systems and to learning a continuum description of cell dynamics in epithelial tissues. We also infer from stochastic particle trajectories the latent phoretic fields driving chemotaxis.
This demonstrates that statistical learning theory combined with physical priors can enable discovery of multi-scale models of non-equilibrium stochastic  processes characteristic of collective movement in living systems.
\end{abstract}

\maketitle


Collective movement is ubiquitous in living systems and is observed at all spatial scales from groups of animals~\cite{couzin2005effective} and shoal of fish~\cite{toner2005hydrodynamics} to the motion of cells within tissues~\cite{riedel2005self} and cytoskeletal molecules within cells~\cite{nedelec1997self, karsenti2008self}. 
Despite their diverse nature, these systems exhibit common emerging properties, including density-dependent transitions to ordered phases \cite{farrell2012pattern}, persistent trajectories \cite{wensink2012meso}, large density fluctuations~\cite{toner2005hydrodynamics}, and spatiotemporal patterning~\cite{schaller2010polar}. 
Naturally then, a fundamental question of collective self-organized motion is how interactions between constituents at the microscopic scale lead to the emergent dynamics at the macroscopic scale.

This question has been addressed by physical theories of self-organized, non-equilibrium active systems both at the microscopic scale of individual constituents~\cite{vicsek1995novel} and at the macroscopic scale in the form of nonlinear hydrodynamic equations~\cite{bar2020self}. 
Explaining the {\em mechanism} of self-organization, however, requires linking the two levels of description in order to, e.g., predict how the microscopic interaction parameters define the hydrodynamic transport coefficients~\cite{grossmann2014vortex} in a multi-scale model~\cite{gao2015multiscale}. This has been addressed by constructing continuum theories based on coarse-graining a microscopic model~\cite{farrell2012pattern, bertin2009hydrodynamic}, by symmetry arguments~\cite{kruse2005generic}, and by non-equilibrium thermodynamics close to equilibrium~\cite{julicher2018hydrodynamic}. These approaches, however, require closure assumptions, are restricted to microscopic models that are analytically tractable, and tend to produce \textit{over-complete} continuum models that are difficult to understand.

\begin{figure}[!t]
\centering
\includegraphics[width=2.7in]{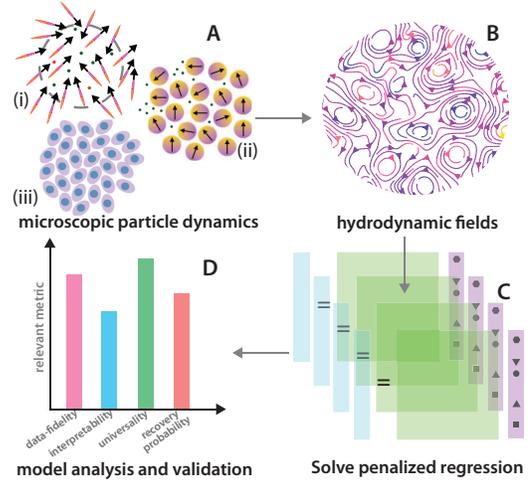}
\caption{\small{Principle of data-driven coarse-graining: {\bf A}: Microscopic data from stochastic active particle (agent) systems. 
{\bf B}: Mean-field quantities of interest computed from particle trajectory data. 
{\bf C}: Statistical learning by solving a penalized regression problem using robustness to data perturbations for model selection. 
{\bf D}: Validation of the \textit{learned} hydrodynamic models using criteria like data fidelity, interpretability, universality, or (statistical) recovery probability.
}}
\label{fig:tell_figure}
\vspace{-2em}
\end{figure}

Analytical coarse-graining is therefore increasingly complemented with data-driven approaches using machine learning~\cite{supekar2021learning}. Statistical learning frameworks have been used to infer effective dynamics from data both in space and time~\cite{rudy2017data, both2021deepmod, brunton2016discovering}, and it has recently been shown how to enforce consistency of the learned models with first principles~\cite{maddu2021learning}. So far, however, most data-driven approaches are applied on one scale, for example to learn particle interaction potentials from particle distributions~\cite{helmuth2010beyond}, to learn force fields and non-equilibrium currents from stochastic trajectories~\cite{frishman2020learning} and Brownian movies~\cite{gnesotto2020learning}, or to learn hydrodynamic equations from hydrodynamic fields~\cite{reinbold2021robust}. 
In contrast to these single-scale applications, data-driven coarse-graining can be used to learn hydrodynamic transport coefficients from microscopic data, to verify closures in kinetic approaches, and to reveal novel principles of self-organization and collective behavior. This has been impressively demonstrated using specialized model designs~\cite{supekar2021learning, 10.7554/eLife.68679}, but a generic framework guaranteeing physical consistency and statistical robustness of the learned models has so far been lacking. 

Here, we provide a principled statistical learning framework for data-driven coarse-graining under a wide class of physical priors. In doing so, we extend the concept of group sparsity~\cite{maddu2021learning} to cross-scale problems. This enables us to learn coarse-grained equations with spatiotemporally varying coefficients, extract the latent stochastic drift fields from particle trajectories, and infer hidden dependences intrinsic to the microscopic system.

\textit{Problem statement}: Given observed trajectories $\bm{r}_p(t_i)$ of active particles (or agents) $p=1,\ldots ,N_p$ along with properties like particle velocity $\bm{v}_p(t_i)$ or orientation $\theta_p(t_i)$ at $\text{T}$ discrete times $t_i = \{t_0, t_0 +\Delta t,\ldots, t_0+\text{T}  \Delta t \}$, infer the functional form of a partial differential equation for hydrodynamic variables of interest (Fig.~\ref{fig:tell_figure}A,B).
Assuming that slowly varying hydrodynamic quantities can be extracted from particle trajectories by averaging, the first step of equation inference is to construct an over-complete {\em dictionary} of all possible right-hand-side terms and numerically evaluate their values on the data~\cite{rudy2017data, brunton2016discovering}.
The canonical form of a one-dimensional model with a single scalar state variable $u$ and a dictionary $\bm{\Theta}\in \mathbb{R}^{\text{N} \times \text{P}}$ of $\text{P} \in\mathbb{N}$ possible terms is:
\begin{equation}\label{eq:explicit_govern}
\underbrace{\begin{bmatrix}
           \vert \\
           u_t \\
           \vert
\end{bmatrix}}_\text{$ \bm{U}_t \in \mathbb{R}^{ \text{N}\times 1}$} 
=
\underbrace{\begin{bmatrix}
          \:\: \vert \qquad \vert \qquad \vert \qquad \vert \qquad \vert \qquad \vert \quad\\
          u\quad uu_{x} \:  \:\:\: \ldots   u_{xx}  \: \ldots \:\:\: \ldots \\
          \:\: \vert \qquad \vert \qquad \vert \qquad \vert \qquad \vert \qquad \vert \quad
\end{bmatrix}}_\text{$\bm{\Theta}\in\mathbb{R}^{ \text{N}\times \text{P}}$}
 \underbrace{ \bm{\xi}}_{\in\mathbb{R}^{\text{P} \times 1}}.
\end{equation}
The left-hand-side vector $\bm{U}_t$ contains the discrete approximations of the time derivatives of $u$ at $\text{N}$ selected space and time points, and each column of $\bm{\Theta}$ contains the discrete approximations of one possible term of the right-hand side at the same data points. 
The problem then amounts to finding a sparse (ideally the sparsest) vector $\bm{\xi}$ such that this model explains the data (Fig.~\ref{fig:tell_figure}C). The sparsest $\bm{\xi}$ defines the simplest sufficient model with the fewest right-hand-side terms. Physical priors, e.g., about symmetries or conservation laws are incorporated by grouping columns of  $\bm{\Theta}$~\cite{maddu2021learning}. 
Thus, the goal is to solve the optimization problem:
\begin{align}\label{groupform}
\hat{\bm{\xi}}^{\lambda} = \arg\min_{\bm{\xi}} &\frac{1}{2} \Vert \bm{U}_t - \sum_{j=1}^{m}  \mathbf{\Theta}_{g_j} \bm{\xi}_{g_j} \Vert_{2}^{2} + \notag \\ 
&\lambda  \sum_{j=1}^{m} \sqrt{p_{g_j}} \: \mathbb{1}  \left ( \Vert \bm{\xi}_{g_j} \Vert_{2} \neq 0 \right),
\end{align}
where $m$ is the number of groups, $\bm{\Theta}_{g_j} \in \mathbb{R}^{\text{N} \times p_{g_j}}$ is the sub-matrix of $\bm{\Theta}$ formed by all columns belonging to group $g_j \subseteq \{1,\ldots,\text{P}\}$, $\bm{\xi}_{g_j} = \{ \bm{\xi}_i: i \in g_j \}$ is the vector $\bm{\xi}$ restricted to the index set $g_j$ of size $p_{g_j}$, i.e., $\vert g_j \vert = p_{g_j}$, and $\mathbb{1}(\cdot)$ is the indicator function. We solve this optimization problem using the gIHT algorithm~\cite{maddu2021learning} based on approximate proximal operators.

\textit{Stability selection chooses the regularization:} Meaningful models are only found if the regularization coefficient $\lambda$ is well chosen~\cite{rudy2017data, both2021deepmod, brunton2016discovering}. We choose it automatically so as to maximize the stability of the inference, i.e., to favor models that are robust to random variations in the data~\cite{maddu2021learning}. The probability that group $g_j$ is selected (i.e., the corresponding entries in $\bm{\xi}$ are non-zero) at a given value of $\lambda$ is $\widehat{\Pi}_{j}^{\lambda} = B^{-1} \sum_{b=1}^{B} \mathbb{1} (g_j \cap \hat{S}^{\lambda}[I_{b}^{*}] \neq \emptyset)$ with $I^{*}$ an independent random sub-sample of the data of size $\vert I^{*} \vert = \lfloor \text{N}/2 \rfloor$ and $\hat{S}^{\lambda}[I_{b}^{*}] = \{ g_j: \Vert \bm{\xi}_{g_j}^{\lambda} \Vert_2 \neq 0 \}$ the inferred group coefficients for this sub-sample. This process is independently repeated for $B\in\mathbb{N}>1$ $(\approx 100)$
different sub-samples of the data. Using the statistical theory of stability selection~\cite{meinshausen2010stability} the stable groups can be extracted as the set $\mathbf{S} (\lambda)= \{ {g_j}: \widehat{\Pi}_{j}^{\lambda} \geq \pi_\text{th}\}$ with the threshold probability $\pi_\text{th}$ chosen to control the number of expected false positives, $E_\text{fp}$, corresponding to spurious right-hand-side terms~\cite{meinshausen2010stability, maddu2021learning}; see Supplement for details. For every identified model $\mathbf{S}^{*}$, we also compute its {\em recovery probability} $\mathrm{P}(\exists  \lambda:\pm\mathbf{S}(\lambda)  = \pm \mathbf{S}^{\ast})$ for increasing sample size $\text{N}$.

\begin{figure}[!t]
\centering
\includegraphics[width=3.2in]{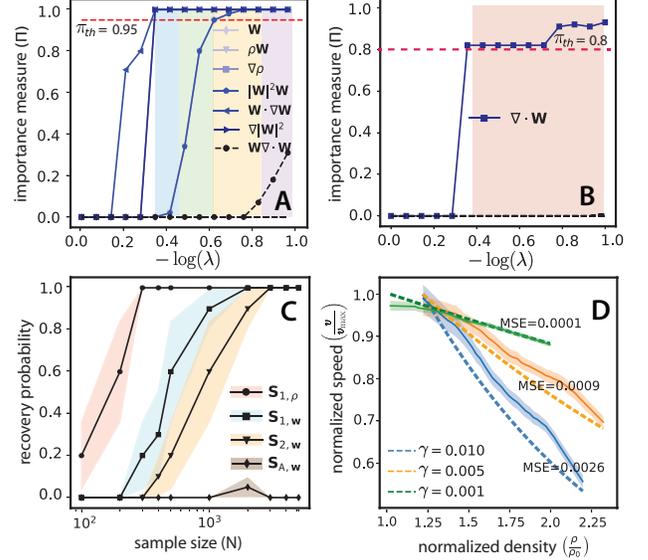}
\caption{\small{Learning hydrodynamic equations for the SPP model. {\bf A}: Stability plot for learning polarity dynamics with dictionary ($\Theta$) size $\text{N}=3000$, $\text{P}=14$. The blue and green regions indicate the regularization ranges where the model  $\mathbf{S}_{1,\mathbf{w}}$ is recovered with $E_\text{fp} \leq 1$ and $E_\text{fp} \leq 2$, respectively. In the yellow region, the model $ \mathbf{S} _{2,\mathbf{w}} $ is recovered with $E_\text{fp} \leq 3$ and in the violet region with $E_\text{fp} > 3$. In all cases, the threshold is $\pi_\text{th} = 0.95$ (red dashed line). Line markers identify model terms as given in the inset legend. {\bf B}: Stability plot for learning density dynamics with $\text{N}=3000$, $\text{P} = 10$. In the colored regularization range, the model $\mathbf{S}_{1,\rho}$ is recovered with $E_\text{fp} \leq 1$ for $\pi_\text{th} = 0.8$, and the correct average lane speed ($\approx 2.1$) is estimated.
{\bf C}: Achievability plot to gauge the identifiability of the models $\mathbf{S}_{1,\rho}$, $\mathbf{S}_{1,\mathbf{w}}$, $\mathbf{S}_{2,\mathbf{w}}, \mathbf{S}_{A,\mathbf{w}}$ for increasing sample size $\text{N}$. 
Colored bands are Bernoulli standard deviations for 20 independent trials. {\bf D}: Inferred latent density dependence of the particle speed $v_p$ for varying $\gamma$ and initial density $\rho_0 = 30$. The dotted lines show the true relation $v_p(\rho) = v_0 \mathrm{e}^{-\gamma \rho \pi } + v_1$.
}}
\label{fig:results_SPP}
\end{figure}

\textit{Application to the self-propelled particles (SPP) model:} We consider the self-propelled stochastic particle system with simple alignment interactions \cite{farrell2012pattern}:
\vspace{-1em}
\begin{equation}\label{eq:SPmodel}
\dot{\bm{r}}_p = v_p \bm{e}_{\theta_p};  \:\dot{\theta}_p = \beta \sum_{q=1}^{N_p} F(\theta_q - \theta_p, r_{qp} ) + \sqrt{2 \epsilon} \,\eta_p(t),
\vspace{-1em}
\end{equation}
where $ \bm{e}_{\theta_p} = \{ \cos \theta, \sin \theta \}$ is the alignment vector of particle $p$ moving at speed $v_p=\| \bm{v}_p \|_2$, and $r_{pq} = \| \bm{r}_q - \bm{r}_p\|_2$. The parameters $\beta$ and $\epsilon$ describe the alignment and fluctuation strengths, respectively; $\eta_p(t)$ is a Gaussian white noise with zero mean and unit variance. The particle alignment function has the form $F(\theta, r_{pq}) = \sin(\theta)/\pi R^2$ if $r_{pq}  < R$ and $0$ otherwise. Density dependent motility is introduced by making the particle speed $v_p$ depend on the local density $\rho$ as $v_p(\rho) = v_0 \mathrm{e}^{-\gamma \rho \pi} + v_1$, where $v_0 $ and $v_1$ are the speeds in the dilute and crowded limits, respectively. The strength of this dependence is set by the parameter $\gamma$. 
This simple microscopic model is able to generate a variety of patterns, ranging from propagating density lanes to moving particle clumps and aster formation. None of those are observed in the standard Vicsek model~\cite{vicsek1995novel} nor its hydrodynamic limit~\cite{toner2005hydrodynamics, mishra2010fluctuations}.

We start by exploring the SPP model close to the Vicsek limit (here $\gamma = 10^{-5}$), where existing hydrodynamic theories are available to compare with~\cite{ toner2005hydrodynamics}.
Figure~\ref{fig:results_SPP}A shows $\widehat{\Pi}(\lambda)$ for the hydrodynamic polarization density $\mathbf{w}(\bm{r}) = \int f(\bm{r},\theta)  \mathbf{e}_\theta \,\mathrm{d}\theta$ and the density $\rho (\bm{r}) = \int f(\bm{r}, \theta)\,\mathrm{d}\theta$ with particle distribution function $f(\bm{r}, \theta) = \sum_{p=1}^{N_p} \delta (\bm{r} - \bm{r}_p) \delta (\theta - \theta_p)$. Above a model probability threshold of $\pi_\text{th}=0.95$, we observe two models for polarization density $\mathbf{S}_{1,\mathbf{w}}, \mathbf{S}_{2,\mathbf{w}}$ (see Table~\ref{table:SPP}) that are consistent and robust with varying levels of fidelity (shown by different color shades) across different $\lambda$ values.
For the density field (see Fig.~\ref{fig:results_SPP}B), we recover the continuity equation $\partial_t \rho = d_4 \nabla \cdot \mathbf{w}$ as model $\mathbf{S}_{1,\rho}$ with the coefficient $d_4$ approximating the constant particle speed, i.e., $d_4 \approx v$.

\begin{table}
\resizebox{27em}{!}{
\setlength\cellspacetoplimit{4pt}
\setlength\cellspacebottomlimit{4pt}
\textcolor{black}{
\begin{tabular}{ Sc|Sc Sc Sc Sc Sc Sc Sc Sc Sc}
 $\mathbf{S}_{1,\rho}$& $\partial_t \rho =$& $c_4\nabla \cdot \mathbf{w} $  & & & & & & & \\ \hline
     $\mathbf{S}_{1,\mathbf{w}}$ & $\partial_t \mathbf{w}=$ & $d_3\nabla \rho$  & $+d_{12}\mathbf{w} \cdot \nabla \mathbf{w}$  & $+d_{7}\nabla \textrm{w}^2$  &   & &   & & \\ \hline
  $\mathbf{S}_{2,\mathbf{w}}$ & $\partial_t \mathbf{w}=$ & $d_3\nabla \rho$  &  $+d_{12}\mathbf{w} \cdot \nabla \mathbf{w}$ & $+d_{7}\nabla \textrm{w}^2$ &  $+d_2\rho \mathbf{w}$ & $+d_1\mathbf{w}$  &   $+d_4\textrm{w}^2 \mathbf{w}$ & &   \\ \hline
 $\mathbf{S}_{A,\mathbf{w}}$& $\partial_t \mathbf{w}=$ & $d_3\nabla \rho$  & $+d_{12}\mathbf{w} \cdot \nabla \mathbf{w}$  & $+d_7\nabla \textrm{w}^2$ & $+d_2\rho \mathbf{w}$ & $+d_1\mathbf{w}$ &  $+d_4\textrm{w}^2 \mathbf{w}$ & $+d_5\mathbf{w}\nabla \cdot \mathbf{w} $
\end{tabular} 
}
}
\vspace{-0.5em}
\caption{Comparison of hydrodynamic models derived from kinetic theory {($\mathbf{S}_{A,\mathbf{w}},\mathbf{S}_{1,\rho}$)} with models {($\mathbf{S}_{1,\mathbf{w}},\mathbf{S}_{2,\mathbf{w}},\mathbf{S}_{1,\rho}$)} learned from microscopic simulation data of the self-propelled particle system (Eq.~\ref{eq:SPmodel}) in the Vicsek limit $\gamma \ll 1$.}
\label{table:SPP} 
\vspace{-1.5em}
\end{table}

We check the statistical consistency of the learned models by computing their recovery probability across all sparse models. We find that the three models $\mathbf{S}_{1,\mathbf{w}} , \mathbf{S}_{2,\mathbf{w}}, \mathbf{S}_{1,\rho} $ can be learned with recovery probability increasing with sample size $\text{N}$, see Fig.~\ref{fig:results_SPP}C. 
We compare this with the over-complete hydrodynamic model $\mathbf{S}_{A,\mathbf{w}}$ (see Table~\ref{table:SPP}) derived from kinetic theory~\cite{farrell2012pattern}, which cannot be recovered without lowering $\pi_\text{th}$ or including more false positives into the model set. This is because the convective term $\mathbf{w} \nabla \cdot \mathbf{w}$ is statistically exchangeable with other terms in the dictionary, since its dynamics are effectively captured by terms like $\mathbf{w} \cdot \nabla \mathbf{w}$ and $\nabla \vert \mathbf{w}\vert^2$. Therefore, our approach finds a simpler model that is consistent with physical priors and is robustly recoverable from the data. 
 
Numerical simulation of the model $\mathbf{S}_{1,\rho} \! + \mathbf{S}_{2,\mathbf{w}}$ reveals propagating density stripes or ``lanes'' that accurately capture the speed observed in microscopic SPP simulations (see Supplementary Video). The sparser model $\mathbf{S}_{1,\mathbf{w}}$ does not produce stripe patterns in simulations, suggesting a necessity for alignment interactions through $\mathbf{w}$ and $\mathbf{w} \vert \mathbf{w}\vert^2$. 

Finally, we look at higher values of $\gamma$, where crowding effects are not negligible. There, learning hydrodynamic models using generic dictionaries without additional structural priors results in an ill-posed problem, as the coefficients of the underlying hydrodynamic model can  be a function of local particle density. In our framework, however, we can explicitly accommodate for this by grouping dictionary columns based on local density and using a block-diagonal dictionary design. The problem then becomes well posed, and we recover a stable model $\mathbf{S}_{1,\rho} = d_4(\rho) \nabla \cdot \mathbf{w} \vert_{\rho_k}, \: \forall k = 1,\ldots , \vert g \vert$, with a coefficient $d_4(\rho)$ that depends on density (see Supplement). From dimensional analysis, we can interpret $d_4(\rho)$ as the local particle speed depending (nonlinearly) on the density and the crowding parameter $\gamma$ (see Fig.~\ref{fig:results_SPP}D). Therefore, this constitutive relation can be learned automatically from data.
\begin{figure}[!t]
\centering
\includegraphics[width=3.2in]{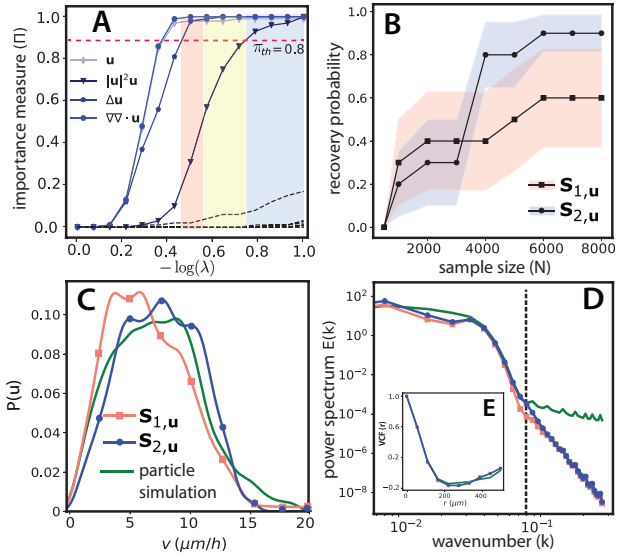}
\caption{\small{Learning hydrodynamic equations from stochastic cellular dynamics. {\bf A}: Stability plot for $\text{N}=5000$, $\text{P}=14$. The orange and yellow regions indicate the regularization ranges where the model $\mathbf{S}_{1, \bm{u}}$ is recovered with $E_\text{fp} \leq 1$ and $E_\text{fp} \leq 2$, respectively. In the blue region, the model $\mathbf{S}_{2,\bm{u}}$ recovered with $E_\text{fp} \leq 2$. Line markers identify model terms as given in the inset legend. {\bf B}: Achievability plot for the models $\mathbf{S}_{1, \bm{u}}, \mathbf{S}_{2, \bm{u}}$ for increasing sample size $\text{N}$. 
Colored bands are Bernoulli standard deviations for 20 independent trials. 
{\bf C}: Comparison of the velocity magnitude $\vert \bm{u}\vert$ distribution from both learned hydrodynamic models and the stochastic cellular simulation. {\bf D, E}: Same comparison for the velocity power spectrum and the velocity correlation function (VCF). 
The vertical dashed line indicates the wavenumber (${k}' = 2 \pi / h $) associated with the bandwidth $h$ of the Gaussian coarse-graining  kernel.}}
\label{cell_part_results}
\end{figure}

\textit{Learning continuum models of collective cell dynamics:} Collective cell migration is a hallmark of morphogenetic events in living tissues, for example during embryogenesis, wound healing, and cancer invasion~\cite{alert2020physical}. 
We consider a stochastic model that describes cells as particles moving with velocity $\bm{v}_p$~\cite{deforet2014emergence, sepulveda2013collective} in a thin epithelial tissue sheet, as governed by the Langevin-like process:
\begin{equation}\label{eq:cellpart}
\frac{d\bm{v}_p}{dt} = -\alpha \bm{v}_p + \sum_{q \in \mathcal{N}(p)} \Big [  \frac{\beta}{n_p} (\bm{v}_q - \bm{v}_p) + \bm{f}_{pq} \Big ] + \sigma(\rho_p) \bm{\eta}_p
\end{equation}
for $p=1,\ldots ,N_p$. Each particle $p$ has $n_p$ neighbors within a neighborhood $\mathcal{N}(p)$. The pair-wise forces between particles are modeled through $\bm{f}_{pq}$, and $\alpha, \beta$ control the substrate friction and velocity alignment, respectively.
Cell motion is driven by a noise term of magnitude $\sigma(\rho)$, where $\rho$ is the particle density, generated from an Ornstein-Uhlenbeck process $\bm{\eta}_p$ with correlation time $\tau$; see Supplement for details. 
Since we are interested in learning a mean-field hydrodynamic model of the process, we consider the regime of low density fluctuations. We want to find a model for the coarse-grained velocity field $\bm{u}(\bm{r}) = \frac{  \overline{v} (\bm{r}) }{\rho(\bm{r})} \int f(\bm{r},\theta) \mathbf{e}_\theta \,\mathrm{d}\theta$, where $\overline{v}(\bm{r})$ is the speed at position $\bm{r}$ interpolated from the closest particles.

\begin{table}
\setlength\cellspacetoplimit{4pt}
\setlength\cellspacebottomlimit{4pt}
\begin{tabular}{ Sc|ScScScScSc }
     $\mathbf{S}_{1,\bm{u}}$ & $\partial_t \bm{u} = $ &$d_1\bm{u} $   &  $+d_8\Delta \bm{u}$ & $+d_6\nabla (\nabla \cdot \bm{u})$ & \\ \hline
  $\mathbf{S}_{2,\bm{u}}$ &  $\partial_t \bm{u} = $ &$d_1\bm{u}$ & $+d_8\Delta \bm{u}$ & $+d_6\nabla (\nabla \cdot \bm{u})$  & $+d_4\bm{u} \vert \bm{u}\vert^2$
\end{tabular} 
\center
\vspace{-1.5em}
\caption{Minimal hydrodynamic models learned for the coarse-grained velocity field $\bm{u}$ from simulation data of stochastic cellular dynamics (Eq.~\ref{eq:cellpart}) in epithelial tissue sheets.}
\label{table:cellpart} 
\vspace{-1em}
\end{table}

The inference results in Fig.~\ref{cell_part_results}A suggest the existence of two stable models $\mathbf{S}_{1, \bm{u}}, \mathbf{S}_{2,\bm{u}}$ (see Table \ref{table:cellpart}) that can be recovered with high statistical robustness. 
The model $\mathbf{S}_{2,\bm{u}}$ has better recovery probability with increasing sample size $\text{N}$, as shown in Fig.~\ref{cell_part_results}B. This model lacks the damping term $ \bm{u}\vert \bm{u} \vert^2$.
Analyzing the models, the terms $\bm{u}$ and $\bm{u} \vert \bm{u}\vert^2$ control the order--disorder transition of the system and set the magnitude of the velocity $\bm{u}$. The terms $\Delta \bm{u}$ and $\nabla (\nabla \cdot \bm{u})$ can be interpreted as the active counterparts of the bend and splay moduli from nemato-hydrodynamics~\cite{de1993physics}. 
They microscopically originate from relative alignment interactions in the particle system of Eq.~\ref{eq:cellpart}. 

We numerically validate the sparse models $\mathbf{S}_{1,\bm{u}}, \mathbf{S}_{2,\bm{u}}$ and find  quantitative agreement between the hydrodynamic models and the stochastic microscopic simulations. The algorithmically inferred hydrodynamic equations are able to capture the velocity magnitude distribution (Fig.~\ref{cell_part_results}C), power spectrum (Fig.~\ref{cell_part_results}D), and the velocity correlation function $\textrm{VCF} (\vert \bm{r} - \bm{r}'\vert)= \big \langle \bm{u}(\bm{r},t) \bm{u}(\bm{r}',t)\big \rangle_t$ (Fig.~\ref{cell_part_results}E) of the stochastic particle dynamics. We found the model $\mathbf{S}_{2,\bm{u}}$ to be more numerical stable than $\mathbf{S}_{1,\bm{u}}$ due to the presence of damping term $\bm{u} \vert \bm{u}\vert^2$ that prevents velocity magnitude from exploding.

\textit{Learning hydrodynamic equations of non-equilibrium particle dynamics driven by phoretic fields:} We consider the Phoretic Brownian Particle (PBP) model~\cite{liebchen2017phoretic, liebchen2018synthetic} with self-propelled particles driven by an imposed chemotactic concentration field $c(\bm{r})$, i.e.
\begin{equation}\label{eq:PBP}
    \dot{\bm{r}}_p =  \mathbf{e}_{\theta_p}; \quad \dot{\theta}_p = \beta \mathbf{e}_{\theta_p} \times \nabla c(\bm{r}_p) + \sqrt{2} \,\eta_p(t),
\end{equation}
for $p=1,\ldots ,N_p$. For $\beta > 0$, the active particles turn towards the phoretic gradients, for $\beta < 0$ they turn against the gradient. This alignment interaction is countered by rotational Brownian motion with Gaussian white noise $\eta_p$ with zero mean and unit variance. 
Contrary to the two previous examples, activity in the PBP model is governed by the imposed phoretic field and is devoid of direct particle interactions. 

We challenge our framework to recover a statistically consistent sparse hydrodynamic model given only the positions $\bm{r}_p$ and orientations $\theta_p$ of the particles, but withholding the underlying phoretic field $c(\bm{r})$. Using symmetry arguments, we construct block diagonal dictionaries that can account for a missing latent scalar or vector field (see Supplement).
For the density field $\rho$, we recover the term $\nabla \cdot \mathbf{w}$ with an estimated coefficient $(c_4 \approx 1)$ approximating the right self-advection speed of the particles. 
For the polarization density $\mathbf{w}$, we consistently identify the model $\mathbf{S}_{1,\mathbf{w}}$ given in Table~\ref{table:chemotaxis}. The estimated phoretic gradient field closely approximates the true one (see Fig.~\ref{fig_driftfield}A). 
The estimate is obtained by smoothness-constrained least-squares regression; see Supplement.

In Fig.~\ref{fig_driftfield}B--D, we compare the learned model $\mathbf{S}_{1,\mathbf{w}}$ with the model $\mathbf{S}_{A,\mathbf{w}}$ (see Table~\ref{table:chemotaxis}) derived from kinetic theory~\cite{liebchen2017phoretic}, and with direct PBP simulations. The agreement suggests that for the particular parameter values chosen, the simpler model $\mathbf{S}_{1,\mathbf{w}}$ is sufficient to describe the hydrodynamics of the PBP system.
As shown in the Supplement, the model $\mathbf{S}_{1,\mathbf{w}}$ can also be derived from kinetic theory under the assumption of low anisotropy.
Thus, at low anisotropy (i.e., small drift, small phoretic gradient), it is not surprising that the model $\mathbf{S}_{1,\mathbf{w}}$ is sufficient to capture the coarse-grained dynamics.

\begin{table}
\resizebox{27em}{!}{
\setlength\cellspacetoplimit{4pt}
\setlength\cellspacebottomlimit{4pt}
\begin{tabular}{ Sc|ScScScScScScScSc }
 $\mathbf{S}_{1, \rho}$ & $\partial_t \rho = $ & $c_4 \nabla \cdot \mathbf{w} $ &&&& \\
  $\mathbf{S}_{1,\mathbf{w}}$ & $\partial_t \mathbf{w} = $ & $d_1\mathbf{w}$ & $+ d_3\nabla \rho $ & $+\rho \nabla c$ &    &  &   \\ \hline
 $\mathbf{S}_{A, \mathbf{w}}$ &  $\partial_t \mathbf{w} = $ &$d_1\mathbf{w}$ & $+d_3\nabla \rho$ & $+\rho \nabla c$ &  $+d_7\nabla \textrm{w}^2$ & $+(\nabla \mathbf{w})^{\top} \cdot \nabla c$ & $+\nabla c \cdot \nabla \mathbf{w}$ & $+(\nabla \cdot \mathbf{w}) \nabla c$ 
\end{tabular} 
}
\vspace{-0.5em}
\caption{Hydrodynamic models of the PBP system (Eq.~\ref{eq:PBP}) learned from microscopic simulation data ($\mathbf{S}_{1,\rho}$, $\mathbf{S}_{1,\mathbf{w}}$) and derived from kinetic theory ($\mathbf{S}_{A,\mathbf{w}}$).}
\label{table:chemotaxis} 
\vspace{-1.0em}
\end{table}

\begin{figure}[!t]
\centering
\includegraphics[width=3.25in]{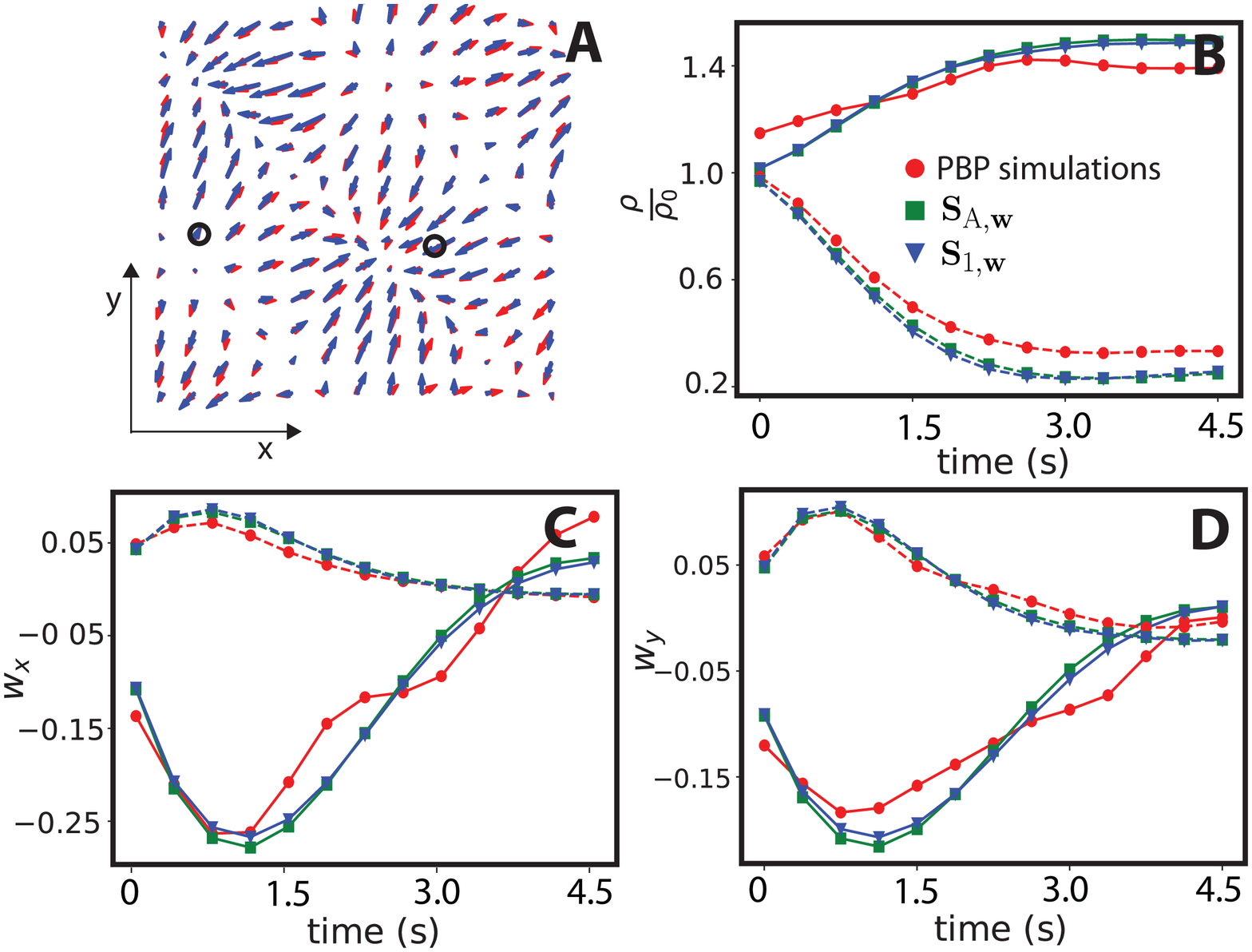}
\caption{\small{Learning hydrodynamic equations for the PBP model with $\mathrm{Pe} = 1, \beta = 0.3$. {\bf A}: Smoothness-constrained regression estimate of the latent phoretic field $\nabla c$ in the simulation domain $[0, 2\pi] \times [0, 2\pi]$. The ground truth $\nabla c$ (red) compared with the inferred phoretic field $\widetilde{\nabla c}$ (blue).
The Relative Mean Squared Error $\left( \frac{\Vert \nabla c - \widetilde{\nabla c} \Vert_2^2}{\Vert \nabla c \Vert_2^2}\right)$ is $ \text{RMSE} = 0.2654$. {\bf B--D}: Time trajectories from the learned ($\mathbf{S}_{1,\mathbf{w}}$) and analytical ($\mathbf{S}_{A,\mathbf{w}}$) models of the density $\rho$ and both components of the polarization density $\mathbf{w}$ at two random points in the spatial domain with coordinates $(0.6, 2.7)$ (dashed) and $(4.4, 2.5)$ (solid), highlighted by circle symbols in (A), compared with those from direct microscopic PBP simulations.}}
\label{fig_driftfield}
\vspace{-1.5em}
\end{figure}

\textit{Conclusions:} We have presented a principled data-driven strategy for automated learning of hydrodynamic equations from stochastic simulations of non-equilibrium active particle dynamics. We inferred sparse hydrodynamic models from simulation data of well-studied SPP and PBP models and found agreement with analytical models derived by kinetic approaches. Numerical simulations of the learned models reproduced relevant metrics of the stochastic particle dynamics, providing quantitative validation. Importantly, our strategy allows for encoding prior knowledge about the system via group sparsity. We have shown that this enables recovery of hydrodynamic equations with spatially or temporally varying coefficients and inference of spatially varying latent fields. 

Beyond benchmarking with existing analytical models, we applied our framework to a stochastic non-equilibrium model of cell motility in living tissues, where coarse-graining via kinetic approaches is not feasible due to the analytically intractable 
particle interactions. Our data-driven framework found two simple models with interpretable terms and long length- and time-scale spectral characteristics in agreement with the stochastic microscopic dynamics. For the PBP system, we used the learned models to identify a sufficient closure for the kinetic approach to quantitatively reproduce the particle dynamics. 

Future work could incorporate stochastic force inference techniques~\cite{frishman2020learning, gnesotto2020learning} into our framework in order to learn hydrodynamic equations from experimental observations of non-equilibrium particle dynamics.

\begin{acknowledgments}
This work was supported by the German Research Foundation (Deutsche Forschungsgemeinschaft, DFG) under Germany’s Excellence Strategy -- EXC-2068-390729961 -- Cluster of Excellence ``Physics of Life'' of TU Dresden, and by the Center for Scalable Data Analytics and Artificial Intelligence (ScaDS.AI) Dresden/Leipzig, funded by the Federal Ministry of Education and Research (Bundesministerium f\"{u}r Bildung und Forschung, BMBF).
\end{acknowledgments}

\bibliography{apssamp}

\appendix

\clearpage
\onecolumngrid

\end{document}


\title{Supplementary information}
\maketitle

\section{Dictionary construction}

In this section, we describe the procedure to extract relevant hydrodynamic variables like density, polarity, and velocity from stochastic particle simulation data. We employ the method of adaptive two-stage estimation \cite{abramson1982bandwidth} to evaluate coarse-grained density field estimates $\widetilde{\rho}(\bm{r})$ from the particle positions $\bm{r}_p, \forall p = {1,...,N_p}$. We start by determining the local bandwidth factor $\Delta_p$ for each particle as follows,
\begin{equation}\label{eq:local_factor}
\Delta_p =  \Bigg \{ \frac{\widetilde{f}(\bm{r}_p)}{G} \Bigg\}^{-\alpha}, \: \forall p={1,...,N_p},
\end{equation}
with the conventional choice of $\alpha=0.5$, and $G$ being the geometric mean of the pilot density estimate $\widetilde{f}(\bm{r})$ \cite{van2003adaptive}. The pilot density estimate $\widetilde{f}(\bm{r})$ is computed from a standard fixed bandwidth $l_\mathbf{K}$ chosen manually, i.e. 
\begin{equation}
\widetilde{f}(\bm{r}) = \sum_{p} \mathbf{K}(\bm{r} - \bm{r}_p, \Sigma ), \: \textrm{ with } \mathbf{K}(\bm{x}, \Sigma ) = \frac{1}{ (\sqrt{2 \pi})^\text{dim}  \vert \Sigma \vert^{1/2}} \exp \left(- \frac{1}{2} \bm{x}^T \Sigma^{-1} \bm{x}\right),
\end{equation}
and $\Sigma=\text{diag}(l_\mathbf{K}^2, l_\mathbf{K}^2)$ is the covariance matrix and $\text{dim}=2$ for all our cases. Once the local bandwidth factors $\Delta_p$ for each particle have been determined using Eq.~\ref{eq:local_factor}, the adaptive kernel density estimate can be computed at a desired location $\bm{r}$ as follows,
\begin{equation}\label{eq:density_kernel}
\vspace{-0.5em}
\widetilde{\rho}(\bm{r}) = \sum_{p} \mathbf{K} (\bm{r} - \bm{r}_p, \Sigma_p), 
\vspace{-0.5em}
\end{equation}
with $\Sigma_p = \text{diag}\left( \Delta_p^2 l_\mathbf{K}^2, \Delta_p^2 l_\mathbf{K}^2 \right)$ is the adaptive covariance matrix associated with each particle. Similarly, we can also compute the coarse-grained estimate of a vector field from the particle properties like orientation and velocity as,
\begin{equation}\label{eq:vel_kernel}
\vspace{-0.5em}
\widetilde{\bm{u}}(\bm{r}) = \sum_{p}  \mathbf{K} (\bm{r} - \bm{r}_p, \Sigma_p) \bm{u}_p.
\vspace{-0.5em}
\end{equation}
$\bm{u}_p$ is the vector-valued particle property. To simplify the notations, we drop the $\sim$ convention and from now refer to the estimates $\widetilde{\rho} = \rho$ and $ \widetilde{\bm{u}} = \bm{u}$.

The hydrodynamic quantities computed in space and time are then used to construct the candidate library $\mathbf{\Theta}$. The library is usually arbitrary with inclusion of all combinations of spatial derivatives and non-linearities up-to a certain order computed for the state variable as discussed in \cite{rudy2017data, brunton2016discovering, maddu2021learning}. However, one can further constrain the solution space of the models using arguments of broken mirror symmetry \cite{supekar2021learning} or constrains informed by domain knowledge \cite{reinbold2021robust}.
In our study, for the scalar variable library construction, it suffices to consider candidate terms that can be written as the divergence of some flux contribution and polynomials of scalar variables. For instance, for learning dynamical equation for the scalar density field $\rho$ which also contains an additional vector valued state variable $\bm{u}$, we consider the candidate library as follows,
\begin{align}\label{eq:scalar_library}
\partial_t \rho =  \: c_0 + c_1 \rho + c_2 \rho^2 + c_3 \vert \bm{u}\vert^2 + c_4 \nabla \cdot \bm{u} + c_5 \nabla \cdot (\rho \bm{u}) + c_6 \Delta \rho  + c_7 \Delta \rho^2 + c_8 \nabla \cdot \left( \vert \bm{u} \vert^2 \bm{u}\right) + c_9 \nabla \cdot (\rho^2 \bm{u}).
\end{align}
Here $c_i, i=0,...,9$ are the coefficients of the equation terms. The library for the vector valued state variable $\bm{u}$ is constructed in a similar fashion as described in \cite{supekar2021learning}, albeit without the contribution of the chiral terms.
\begin{align}\label{eq:vector_library}
\partial_t \bm{u}   = & \: d_0 + d_1 \bm{u} + d_2 \rho \bm{u} + d_3 \nabla \rho + d_4 \vert \bm{u}\vert^2 \bm{u} + d_5 \bm{u} \nabla \cdot \bm{u} 
 + d_6 \nabla \nabla \cdot \bm{u} + d_7 \nabla \vert \bm{u} \vert^2 + d_8 \Delta \bm{u} + d_9 \Delta^2 \bm{u} + d_{10} \vert \bm{u}\vert^2 \nabla \rho   \nonumber \\ 
& + d_{11} \nabla \rho \cdot \nabla \bm{u} + d_{12} \bm{u} \cdot \nabla \bm{u} + d_{13} \nabla \rho (\nabla \cdot \bm{u}),
\end{align}
Here $d_i, i=0,...,13$ are the coefficients of the equation terms. 
Inclusion of terms like $\Delta^2 \bm{u}$ follows pattern-forming models like Swift-Hohenberg \cite{wensink2012meso, supekar2021learning}. 
We expand the library of the vector state-variable ($\bm{u}$) in Eq.~\ref{eq:vector_library} to account for the unobserved scalar and the vector fields in a way that still obeys the symmetry. For instance, for inferring the mean-field equations of the PBP system, we append to the library few scalar and second order tensor quantities $\{ \rho, \vert \bm{u}\vert^2, \nabla \cdot \bm{u}, \nabla \bm{u}, \nabla \bm{u}^T \}$ to allow for the candidate terms $\{ \rho \nabla c, \vert \bm{u}\vert^2 \nabla c, (\nabla \cdot \bm{u} ) \nabla c, \nabla c \cdot \nabla \bm{u},  \nabla \bm{u}^T \cdot \nabla c \}$ to be included into the model. Here $c$ and $\nabla c$ correspond to the unknown scalar and the vector field, respectively. In section~\ref{sec:priors}, we discuss how block diagonal dictionary design and grouping equation terms can allow for encoding physical priors into the learning problem.

\section{Model selection with stability selection}
\begin{figure}
\centering
\includegraphics[width=7in]{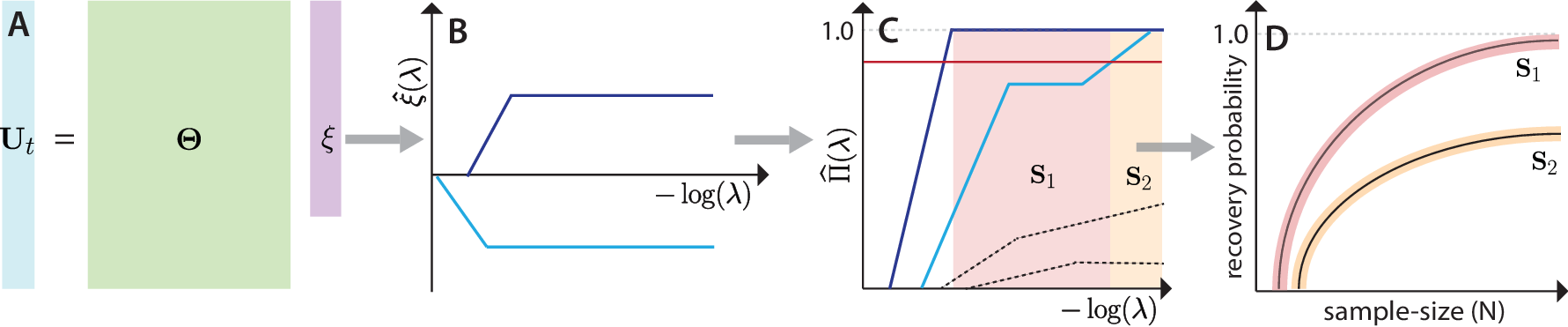}
\caption{\small{\textbf{Schematic for model selection}: (A) Construct the left-hand side vector $\mathbf{U}_t$ and the dictionary $\mathbf{\Theta}$ from coarse-grained particle data. (B) Computation of the regularization path $\widehat{\xi} (\lambda)$ for varying regularization parameter $\Lambda = \{ \lambda_\text{max}, \lambda_\text{min}\}$. The colored lines are the non-zero coefficients appearing along the $\lambda$-path. (C) Evaluate importance measures $\widehat{\Pi}$ for each dictionary component for $\lambda \in \{ \lambda_\text{max}, \lambda_\text{min} \}$. Model selection based on thresholding of importance measure $\widehat{\Pi}(\lambda)$ as per Eq.~\ref{eq:bounds}. The solid lines in blue shades are more robust in comparison to other components shown in dotted black lines. $\mathbf{S}_1$ and $\mathbf{S}_2$ are the supports/models identified in the red and orange region, respectively and $\vert \mathbf{S}_2 \vert \geq \vert \mathbf{S}_1 \vert$ for a chosen $E_\text{fp}$.  (D) Check for the statistical consistency of the identified models $(\mathbf{S}_1, \mathbf{S}_2)$ for increasing sample-size $\text{N}$. }}
\label{fig:explain_statistics}
\end{figure}

Once the dictionary is constructed (Fig.~\ref{fig:explain_statistics}A), we solve the sparse regression problem Eq.~{2} (main text) using the gIHT algorithm \cite{maddu2021learning} for various regularization parameters $\lambda$ that lead to the construction of the {\em regularization path} as illustrated in Fig.~\ref{fig:explain_statistics}B. A typical choice of regularization interval $\Lambda = \{ \lambda_\text{min} , \lambda_\text{max} \}$ with $\lambda_\text{max} = \max_{j \in \{1, \ldots ,m\}} \frac{1}{2} \Vert \bm{\Theta}_{g_j}^{\top} \bm{U}_t \Vert_2^2$ as computed from the Karush-Kuhn-Tucker conditions~\cite{rudy2019data} and $\lambda_\text{min} = 0.01 \lambda_\text{max}$.
For model selection, we rely on the statistical principle of stability selection which computes robustness of each group under random sub-sampling of the data~\cite{meinshausen2010stability}.
We perform stability selection by generating $B$ random
sub-samples $I_b^{*}$, $b=1,\ldots ,B$ of the data and using the gIHT algorithm to find the set $\hat{S}^{\lambda}[I_b^{*}] \subseteq \{1,\ldots,P\}$ of coefficients (or groups) for every data sub-sample $I_b^{*}$ for different values of $\lambda \in \Lambda$. The probability that group $j$ overlaps with the selected subset for a given $\lambda$ is approximately~\cite{meinshausen2010stability}.
\begin{subequations}
\begin{align}
    \hat{\Pi}_{g_j}^{\lambda} &= \mathbb{P}[ g_j \cap  \hat{S}^{\lambda} \neq \emptyset ] \\
                            &\approx \frac{1}{B}\sum_{b=1}^{B} \mathbb{1} (g_j \cap \hat{S}^{\lambda}[I_{b}^{*}] \neq \emptyset), \quad g_j \subseteq \{1,...,P\}. \label{eq:importance_measure}
\end{align}
\end{subequations}
This is the {\em importance measure} for group or coefficient $j$. Plotting this importance measure as a function of $\lambda\in\Lambda$ provides an interpretable way to assess the robustness of the estimation across levels of regularization in a so-called {\em stability plot} as illustrated in Fig.~\ref{fig:explain_statistics}C. For any value of the regularization parameter $\lambda$ a set of stable coefficients (or groups) can be selected, i.e.  $\hat{S}_\text{stable} =\{j:\widehat{\Pi}_{g_j}^{\lambda}>\pi_\text{th} \}$. However, we are interested in recovering a stable support $\hat{S}_\text{stable}$ in a statistically consistent manner with minimal number of false positives included. The theory of stability selection allows for choosing the 
threshold $\pi_\text{th}$ to control the type I error of false positives~\cite{buhlmann2014high},
\begin{equation}\label{eq:bounds}
    \pi_\text{th} = \frac{1}{2} + \frac{\binom q{p_g}^2}{2 \binom P{p_g} E_\text{fp}},
\end{equation}
where $E_\text{fp}$ is the upper bound on the expected number of  false positives, $q = \mathbb{E}(\vert \hat{S}^{\lambda}\vert)$ is the average number of selected variables, and ${p_g}$ is the group size. Once a model support $\mathbf{S}^{\ast}$ has been identified, we compute the probability of successful support recovery $\mathbb{P}(\exists  \lambda:\pm\mathbf{S} (\lambda) = \pm \mathbf{S}^{\ast})$ for increasing sample-size $\text{N}$. This probability helps us to quantify the statistical consistency with which the model support $\mathbf{S}^*$ can be identified with the library $\mathbf{\Theta}$.


\newpage
\section{Encoding Knowledge priors into learning}\label{sec:priors}

\begin{figure}
\centering
\includegraphics[width=6in]{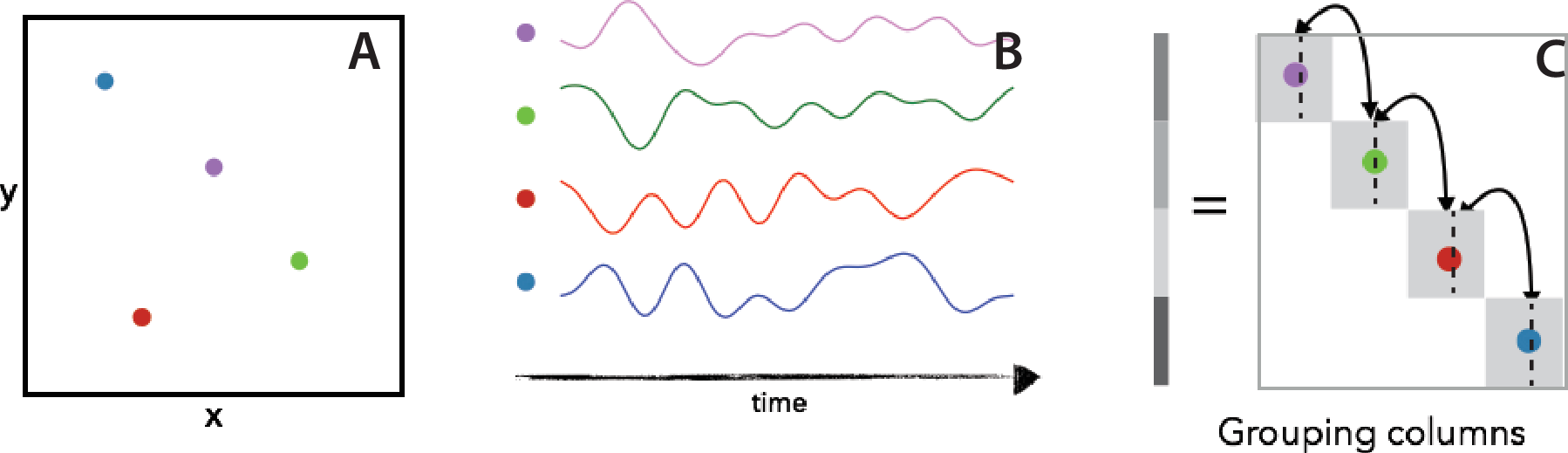}
\caption{\small{\textbf{Encoding knowledge about spatial variability of coefficients using grouping}: Selects points in space (A) and construct dictionary block for each point using the time trajectory information (B). Stack the dictionary blocks in a diagonal fashion and group every column of each block with the corresponding column in the remaining blocks (C) and form the index set as in Eq.~\ref{eq:groupindex}. In (C), all the columns marked by vertical dashed lines form a single group with size $\vert g\vert = 4$.}}
\label{fig:explain}
\end{figure}

We use the idea of grouping columns for enforcing physical priors that enable recovering equations with spatially or temporally varying coefficients. For example, in the SPP system with non-negligible crowding effects, using column grouping we inform the learning algorithm about density dependence of the transport coefficients. To \textit{infer} local density-dependent coefficients, we construct block-diagonal dictionary design $\bm{\Theta} \in \mathbb{R}^{\text{N} \times \text{P}}$ \cite{maddu2021learning} with each dictionary block $\bm{\theta}_b \in \mathbb{R}^{n \times p}$ computed from space and time locations that have approximately the same local density. For the sake of clarity, we distinguish the number $p$ of potential terms in each block and the number $\text{P} = p_b p$ of columns in the overall dictionary. We then can collect these locations into the set $\mathcal{D}_b$ defined as follows,
\begin{equation}
\mathcal{D}_b = \Big \{   \{ \bm{x}, t \} \Big \vert \rho(\bm{x},t) \in \big [\rho_b - \epsilon, \rho_b + \epsilon \big] \Big \},
\end{equation}
with $\bm{x} \in [0, 10] \times [0, 10]$ and $t \in [0, 5]$. In words, the set $\mathcal{D}_b$ consists of all the space and time coordinates that have the same value of density $\rho_b$ up-to a prescribed tolerance $\epsilon$. We then proceed to construct the block library $\bm{\theta}_b$ for various density values with relevant information from the coarse-grained hydrodynamic variables sampled at the space and time locations listed in the set $\mathcal{D}_b$. Finally, the block-diagonal matrix $\bm{\Theta}$ is formed by diagonally stacking $p_b$ number of dictionary blocks $\bm{\theta}_b$.  We then form the groups defined as follows,
\begin{equation}\label{eq:groupindex}
g_l = \big \{ \{l + kp \} \: \forall \: k \in \{0,...,p_b-1 \}\big \}
\end{equation}
Here the set $g_l$ informs the sparse regression algorithm that all the $l^{th}$ columns associated with each dictionary block $\bm{\theta}_b$ are grouped together. Similarly, the same principle can be used to account for spatially or temporally varying latent fields (See Fig.~\ref{fig:explain}).

\subsection{Smoothness constraints on coefficient estimation}\label{smoothness_filter}
The transport coefficients of the inferred hydrodynamic models are directly estimated by the gIHT algorithm. However, for coefficients which have spatial or temporal variability, we consider an additional step of smoothness constrained regression over the identified support as described in \cite{maddu2021learning}. For this, we solve the trend filtering problem,
\begin{equation}
\hat{\bm{\xi}}_s = \arg \min_{\xi} \frac{1}{2} \Big \Vert \bm{U}_t - \sum_{j-1}^{K} \theta_{g_j} \bm{\xi}_{g_j} \Big \Vert_2^2 + \lambda_\text{f} \sum_{j=1}^{K}  \Big \Vert \text{D}_j^{k+1} \bm{\xi}_{g_j} \Big \Vert_1
\end{equation}
where $K$ is the number of distinct coefficient groups identified by the gIHT algorithm. The matrix $\text{D}^{k+1}$ is the discrete smoothing filter based on the $(k+1)th$ derivative. The regularization parameter $\lambda_\text{f}$ controls the smoothness degree imposed on the coefficients. In our study, we use the 2D Laplacian discrete filter, i.e. $k=1$ to impose smoothness on spatially varying coefficients and solve the optimization problem using ADMM algorithm \cite{maddu2021learning}. The regularization parameter of $\lambda_\text{f} = 1$ was used for the phoretic field reconstruction with smoothness constraints (Fig. 4A main text).


\newpage
\section{Simulation details}\label{Simulation details}

In all the agent-based simulations presented in this paper, the temporal dynamics of each agent was computed using the explicit Euler algorithm. A particle neighborhood was constructed using the Verlet list and periodic boundary conditions were employed. All simulations are implemented in the scalable parallel simulation software library OpenFPM~\cite{incardona2019openfpm}. For the numerical solution of the hydrodynamic equations we use fourth-order central differences for spatial discretization with periodic boundary conditions and the predictor-corrector algorithm for time integration. To account for the characteristic flow of information, we use upwind schemes based on second-order ENO differences for spatial discretization whenever required. We list all parameters of both the particle and hydrodynamic models that are used to produce the figures presented in the main text. 

\begin{figure}[!t]
\centering
\includegraphics[width=6in]{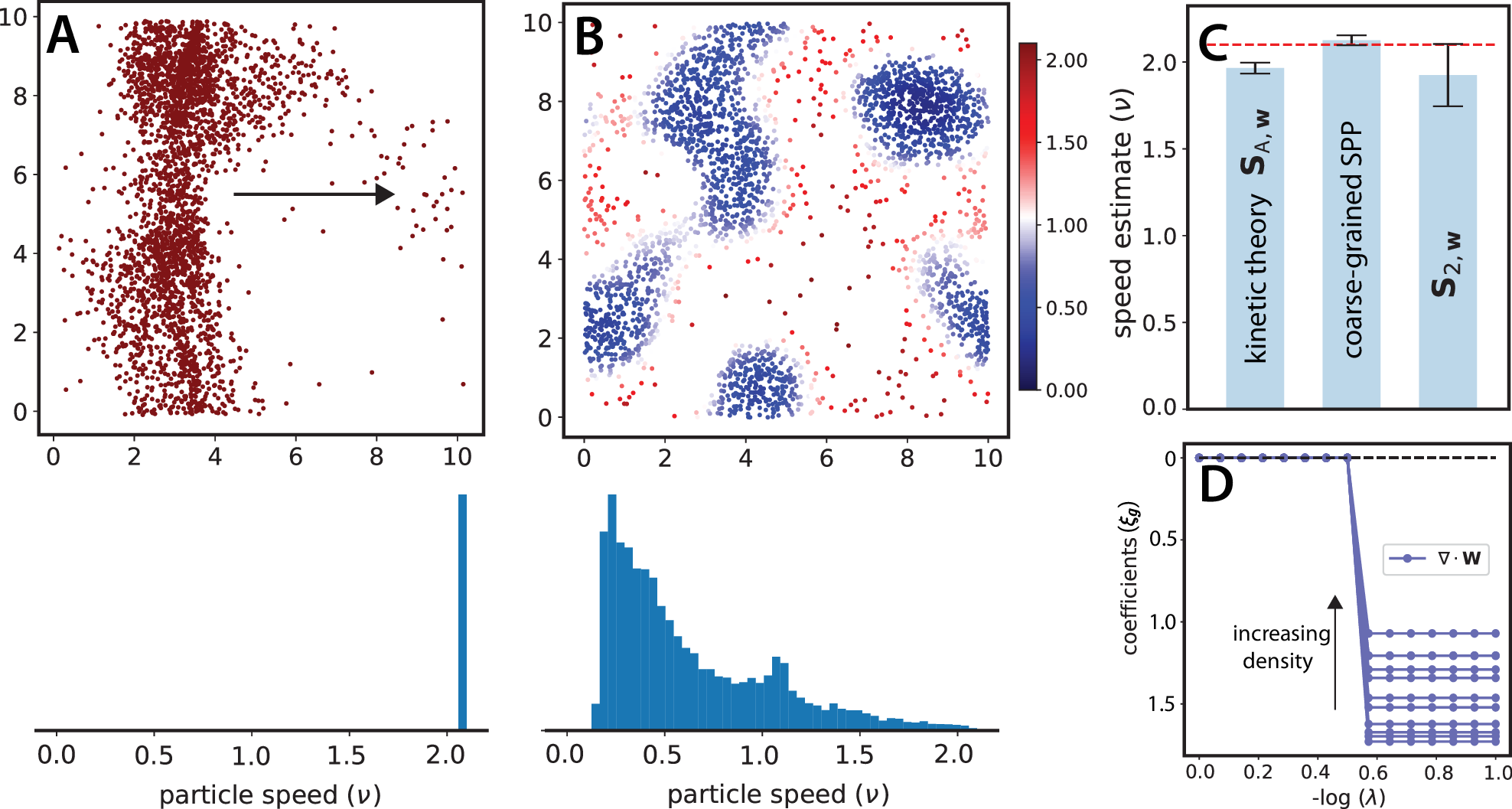}
\caption{\small{\textbf{Spatio-temporal dynamics of the self-propelled particle (SPP) model}: A) Moving stripes emerging in a direction perpendicular to the stripe with no density-dependent motility observed for $\lambda=1e^{-5}, \epsilon=2$. The arrow indicates the direction of motion of the stripe. B) Moving clumps emerging due to the crowding effect for $\lambda=0.0075, \epsilon=2.5$ for N = 3000. Particles are color coded by their respective speeds $\nu$. In the second row we also show the distribution of the particle speed $\nu$. C) Comparison between average particle speed ($\nu$) estimates in the stripe for the kinetic theory ($\mathbf{S}_{\mathbf{A},\mathbf{w}} + \mathbf{S}_{1, \rho}$), SPP simulations and the learned sparse model ($\mathbf{S}_{2,\mathbf{w}} + \mathbf{S}_{1, \rho}$); also see movie in the supplementary. The average particle speed in the stripe computed as $ \nu = v_x/\cos(\delta)$, where $v_x$ is the horizontal speed of the stripe and $\delta$ is the average orientation of the particles in the stripe. The time taken by the stripes to traverse the horizontal spatial axis is used to calculate the horizontal speed $v_x$. D) Typical regularization path $\bm{\xi}_{g_j} (\lambda)$ for learning the density dynamics in the regime of non-negligible crowding effects with block-diagonal dictionary design $(n=200, p = 10, m = 10)$ constructed for SPP data generated with $\gamma = 0.0075$. Marked blue lines correspond to the coefficients ($\bm{\xi}_{g}$) of the columns that belong to the group support: $\{ \nabla \cdot \mathbf{w} \vert_{\rho_k} \} \: \forall \: k = 1,..., \vert g \vert$.}}
\label{fig:phase_viz}
\end{figure}

\textit{Self-propelled particle system}: The SPP model is simulated with $N_p=3000$ particles in a square domain of length $L=10$. The alignment strength and magnitude of fluctuations in Eq.~{3} (main text) are set to $\beta=0.16$ and $\bm{\epsilon}=2$, respectively. The speed in the dilute and crowded limits are $v_0=2$ and $v_1=0.1$, respectively with the radius of interaction kernel $R=1$. For the learning problem, the hydrodynamic quantities density $\rho$ (Eq.~\ref{eq:density_kernel}) and polarization density $\mathbf{w}$ (Eq.~\ref{eq:vel_kernel}) are approximated in the physical domain with $100 \times 100$ uniform spatial resolution using the adaptive kernel density estimator ($l_\mathbf{K}=1.5$ units) and a time resolution $dt=0.1$. 

\begin{table}
\setlength\cellspacetoplimit{4pt}
\setlength\cellspacebottomlimit{4pt}
\begin{tabular}{ Sc|Sc|Sc|Sc|Sc|Sc|Sc|Sc|Sc }
      & $\mathbf{w}$   &  $\rho \mathbf{w} $ & $ \nabla \rho $ & $ \mathbf{w} \cdot \nabla \mathbf{w}$ & $\nabla \vert \mathbf{w}\vert^2$ & $ \mathbf{w}\vert \mathbf{w}\vert^2$ & $\mathbf{w} \nabla \cdot \mathbf{w}$ & $\nabla \cdot \mathbf{w} $\\ \hline
  $\mathbf{S}_{1,\mathbf{w}}$  &  $ \times$& $ \times$ & $  -0.524$ & $-0.0405$ &  $0.0081$ & $\times$ & $ \times$\\ \hline
   $\mathbf{S}_{2,\mathbf{w}}$   & $-0.4523$ & $0.019$ & $ -0.517$ & $-0.0405$ & $0.0081$& -0.00023 &  $ \times$ \\ \hline
      $\mathbf{S}_{1,\rho}$   &  $ \times$ & $ \times$ & $ \times$ & $ \times$ & $ \times$ &  $ \times$  & $ \times$ & -1.0\\ \hline
      $\mathbf{S}_{\mathbf{A},\mathbf{w}}$   & $-0.95$ & $0.038$ & $-0.495$  & $-0.015$ & $0.0125$ & $-0.00076$ & $ -0.025$ \\ \hline
\end{tabular} 
\vspace{-1em}
\center
\caption{The transport coefficients of the sparse hydrodynamic models associated with the SPP model (Eq.~{3} main text) obtained via the gIHT algorithm. The parameters of the analytical model $\mathbf{S}_{A, \mathbf{w}}$ are computed from the kinetic theory \cite{farrell2012pattern}. All the parameters are expressed with unit of length $[\ell] = R$ and time scales measured as $[t] = [\ell]/[v_0 + v_1]$.}
\label{table:SPP_parameters} 
\end{table}

\indent The \textit{learned} hydrodynamic equations are solved in the square domain with length $L=10$ discretized with $100$ computational units in each direction, i.e. $dx = L/100 = 0.1$ with time-step $dt=0.005$. The initial density is taken be same as the SPP simulations, i.e. $\rho(\bm{r},t=0) = 30 $. \\

\textit{Stochastically interacting particle model for cellular dynamics}: The interaction force between two cells $p$ and $q$ is taken as the gradient of the interaction potential $U$, i.e. $\bm{f}_{pq} = -\nabla_p U(\vert \bm{r}_p - \bm{r}_q\vert)$. The potential $U(r)$ has the form,
\begin{equation*}
U(r) = U_0 \exp \left( - \frac{r^2}{a_0^2} \right) + U_1 (r - a_1)^2 H(r - a_1) H(a_m - r) 
\end{equation*}
with parameter values $U_0=2400 \mu m^2/h, a_0 = 8 \mu m, U_1 = 2 \mu m^2/h, a_1 = 35 \mu m, a_m = 80 \mu m $ as in studies  \cite{deforet2014emergence, sepulveda2013collective}, and  $H(x)$ is the Heaviside function with $H(x) = 1$ for $ x > 0$ and $H(x) = 0$ otherwise. The simulation is conducted with $N_p = 4000$ particles uniformly distributed in the square domain with physical dimensions $1000 \mu m \times 1000 \mu m$ which corresponds to the initial cell density $\rho_0 = 4 \times 10^{-3}$ $\mu m^{-2}$. The dissipative parameters are set to $\alpha=1.42 h^{-1}, \beta = 60 h^{-1}$, and the noise amplitude is taken to constant $\sigma = \sigma_0 = 150 \mu m/h^2$. The parameters are chosen as to reproduce velocity statistics of the experiments \cite{deforet2014emergence, sepulveda2013collective}. The stochastic motion of the cells is driven by the noise term $\bm{\eta}_p$ generated from an Ornstein-Uhlenbeck process:
\begin{equation}\label{eq:noise}
\tau \frac{d\bm{\eta}_p}{dt} = -\bm{\eta}_p + \bm{\xi}_p
\end{equation}
with $\bm{\xi}_p$ a delta-correlated white noise drawn at each particle $p$ and the correlation time $\tau = 1.39 h$ \cite{sepulveda2013collective}. We use Euler-Maruyama method with a step-size $dt=0.002 h$ to integrate the Eq.~\ref{eq:noise}. For the learning problem, the coarse-grained velocity $\bm{v}$ is approximated in the physical domain with $100 \mu m \times 100 \mu m$ spatial resolution using the adaptive kernel density estimator ($l_\mathbf{K} = 50 \mu m$ units) and a time resolution $dt=0.002 h$. The \textit{learned} hydrodynamic equations are simulated with parameters listed in Table.~\ref{table:cellpart_parameters}. The initial conditions for the PDE solution is taken directly from the coarse-grained velocity estimates of the particle simulations.\\


\begin{table}
\setlength\cellspacetoplimit{4pt}
\setlength\cellspacebottomlimit{4pt}
\begin{tabular}{ Sc|Sc|Sc|Sc|Sc }
      & $\bm{u}$   &   $\bm{u} \vert \bm{u}\vert^2$ & $\Delta \bm{u}$ & $\nabla \nabla \cdot \bm{u}$ \\ \hline
  $\mathbf{S}_{1,\bm{u}}$  & $2.0$  & & $4.2 \times 10^{3}$ & $-1.9 \times 10^{3}$\\ \hline
   $\mathbf{S}_{2,\bm{u}}$   & $2.7$  & $8 \times 10^{-3}$ & $ 4.2 \times 10^{3} $ & $-1.9 \times 10^{3}$\\ \hline
  units  & $1/h$  & $\mu m^{-2} h$ & $\mu m^2 /h$ & $\mu m^2 /h$\\ \hline
\end{tabular} 
\caption{The transport coefficients of the sparse hydrodynamic models learned from stochastic cellular model (Eq.~4 main text) as estimates of the gIHT algorithm. }
\label{table:cellpart_parameters} 
\end{table}

\textit{Phoretic Brownian particle (PBP) model:} The PBP model (Eq.~{5} in main text ) is simulated with the strength of the phoretic field $\beta = 0.3$. The underlying 2D spatially varying phoretic field $c(x,y)$ is a sinsuoidal forcing term of the form
\begin{equation}\label{Eq:sinfunc}
    c(x,y) = \sum_{i=1}^{M} A_x^i \cos(2\pi l_x^i x/L + \phi_x^i) A_y^i \sin(2\pi l_y^i y/L + \phi_y^i),
\end{equation}
where $M=20$ and $L=2\pi$ is the size of the domain. The parameters $A_x, A_y$ are drawn independently and uniformly from the interval $[-2,2]$, and similarly $\phi_x, \phi_y$ are uniformly drawn from the interval $[0, 2\pi]$. The parameters $l_x, l_y$ which set the local length scales are sampled uniformly from the integer set $\{1,2\}$. The particle simulation is conducted with $\text{N}_p = 5000$ agents for a total time of $t=5$. The coarse-grained density $\rho$ and polarization density $\mathbf{w}$ is computed in the physical domain with $100 \times 100$ spatial resolution using the adaptive kernel density estimator (global bandwidth $l_\mathbf{K}=0.5$ units) for a time resolution $dt=0.0125$.\\
\indent For solution of the hydrodynamic equations, the physical domain $[0, 2\pi] \times [ 0, 2\pi]$ is discretized into $100 \times 100$ grid with periodic boundaries. The equations are solved with the parameters listed in Table.~\ref{table:PBP_parameters}. For the spatially varying pre-factors associated with the imposed phoretic field gradients ($\nabla c$), we use the smoothness constrained regression estimates as described in section \ref{smoothness_filter}.
\begin{table}
\setlength\cellspacetoplimit{4pt}
\setlength\cellspacebottomlimit{4pt}
\begin{tabular}{ Sc|Sc|Sc|Sc|Sc|Sc|Sc|Sc }
      & $\mathbf{w}$   &   $\nabla \rho$ & $\rho $ & $\Delta \mathbf{w}$ & $(\nabla \mathbf{w})^T $ & $ \nabla \mathbf{w}$ & $(\nabla \cdot \mathbf{w})$ \\ \hline
  $\mathbf{S}_{1,\mathbf{w}}$  & $-1.3$  & $-0.48$ & Fig.~$4$A (main text)  &  & & & \\ \hline
 $\mathbf{S}_{A,\mathbf{w}}$  & $-1.0 - \vert \nabla c\vert^2$  & $-0.5$ & $\frac{\beta}{2} \nabla c$ & $-\frac{\text{1}}{16}$  & $(\ast)\cdot \nabla c$ & $\nabla c \cdot (\ast)$ &  $\nabla c$ \\ \hline
\end{tabular} 
\caption{The transport coefficients of the hydrodynamic models associated with the PBP model (Eq.~5 main text) obtained via the gIHT algorithm. The equation were recovered in the non-dimensional form and thus all the units of the pre-factors are dimensionless.}
\label{table:PBP_parameters} 
\end{table}

\begin{figure}[!t]
\centering
\includegraphics[width=6.0in]{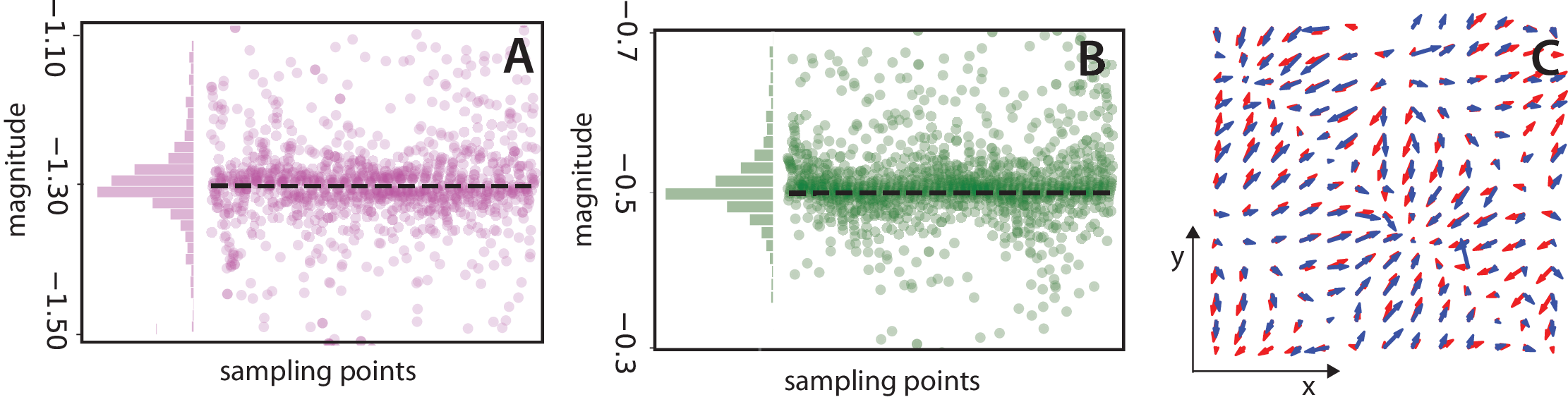}
\caption{\small{\textbf{Group regression parameters of hydrodynamic model of PBP system}: A) Estimate of the prefactor for the $\mathbf{w}$. B) Estimate of the prefactor for the term $\nabla \rho$. C) Estimation of the phoretic field $\nabla c$ without smoothness constraints with $\text{RMSE} = 0.31505$.}}
\label{fig:phase_viz}
\end{figure}
%

\section{Kinetic theory of the PBP model}

Following the derivation of~\cite{liebchen2017phoretic}, the equation of motion for the Fourier modes of the $\text{N}$-particle probability density function $f(\bm{r}, \theta, t)$ for the PBP model is written as:
\begin{equation*}
\dot{f}_k (\bm{r},t) =  - \frac{\textrm{1}}{2} [  \partial_x \left( f_{k+1} + f_{k-1} \right) - \textrm{i}  \partial_y \left( f_{k+1} - f_{k-1} \right) ]  - k^2 f_k + \frac{\beta \vert \nabla c \vert k }{2} \left( f_{k+1} e^{i \delta } - f_{k-1} e^{-i \delta }\right),
\end{equation*}
where $f_k(\bm{r},t) = \int f(\bm{r}, \theta, t)e^{ik\theta}\,\mathrm{d}\theta$ is the $k^\text{th}$ Fourier mode with $f_0 (\bm{x},t) = \rho(\bm{x},t) = \int f(\bm{x},\theta, t)\,\mathrm{d}\theta$ and $\mathbf{w}(\bm{x},t) = (\textrm{Re}f_1, \textrm{Im}f_1) = \int f(\bm{x}, \theta, t) \bm{e}_{\theta}\,\mathrm{d}\theta$. Substituting for $\cos(\delta) = -\partial_x c/\vert \nabla c\vert, \: \sin(\delta) = \partial_y/\vert \nabla c\vert$ and using the closure assumption based on strong penalization for deviation from isotropy, i.e.~$f_2 \approx 0$, the governing equation for the polarization density $\mathbf{w}$ is obtained:
\begin{align*}
\dot{f}_1 (\bm{r},t) & = - \frac{\textrm{1}}{2} [ \partial_x f_0 + i \partial_y f_0  ] - f_1 - \frac{\beta \vert \nabla c \vert }{2} ( f_0 \cos(\delta) - f_0 i \sin(\delta)), \\
\dot{\mathbf{w}} & = - \frac{\nabla \rho}{2} - \mathbf{w} + \frac{\beta \rho \nabla c}{2}.
\end{align*}

\bibliography{apssamp}